\documentclass[twocolumn,aps,prd,showpacs,nofootinbib]{revtex4}

\usepackage{amsmath,amssymb,amsfonts,amsthm}
\bibpunct{}{}{,}{s}{}{,} 
\usepackage{bm}

\newcommand{\be}{\begin{equation}}
\newcommand{\ee}{\end{equation}}
\newcommand{\ba}{\begin{eqnarray}}
\newcommand{\ea}{\end{eqnarray}}
\def\a{\alpha}
\def\g{\gamma}
\def\vp{\varphi}
\def\bu{\bar\mu}
\def\cH{{\cal H}}
\def\cN{{\cal N}}
\def\cV{{\cal V}}
\def\p{\partial}
\newcommand{\Eq}[1]{(\ref{#1})}
\def\lp{\ell_{\rm Pl}}

\begin{document}

\vspace{1cm}

\title{Loop quantum cosmology and tensor perturbations in the early universe}
\author{Gianluca Calcagni}
\email{gianluca@gravity.psu.edu}
\affiliation{Institute for Gravitation and the Cosmos,
Department of Physics,\\ The Pennsylvania State University,
104 Davey Lab, University Park, PA 16802}

\author{Golam Mortuza Hossain}
\email{ghossain@unb.ca}
\affiliation{Department of Mathematics and Statistics,\\ University of New Brunswick, 
Fredericton, NB E3B 5A3, Canada}
\date{\today}
\begin{abstract}
We study the tensor modes of linear metric perturbations within an effective framework of loop quantum cosmology. After a review of inverse-volume and holonomy corrections in the background equations of motion, we solve the
linearized tensor modes equations and extract their spectrum. Ignoring holonomy corrections, the tensor spectrum is
blue tilted in the near-Planckian superinflationary regime and may be observationally disfavoured. However, in this case background dynamics is highly nonperturbative, hence the use of standard perturbative techniques may not be very reliable. On the other hand, in the quasi-classical regime the tensor index receives a small negative quantum correction, slightly enhancing the standard red tilt in slow-roll inflation. We discuss possible interpretations of this correction, which depends on the choice of semiclassical state.
\bigskip

\end{abstract}

\maketitle


\section{Introduction}

One of the major problems of modern theoretical physics is how to reconcile the highly successful framework of quantum field theory with Einstein's general relativity, that describes classical gravitational interaction with astonishing accuracy. There exist several approaches in this direction. One, extensively studied by the scientific community, is {\em string theory}, which attempts to unify all forces of nature. Another, dubbed {\em loop quantum gravity} (LQG), focusses on a consistent quantization of general relativity alone \cite{rov97,thi02,ALReview,smo04,rov04}. While the understanding of physical consequences implied by full LQG is still an open research topic, progress has been made by studying the theory in symmetry-reduced spacetimes. In particular, {\em loop quantum cosmology} (LQC) opens up a possibility of resolving the singularities that plague classical cosmological spacetimes \cite{boj1,boj2,boj3,boj4,ABL,bojLRR}. For a spatially flat Friedmann--Robertson--Walker (FRW) background with a massless scalar field, the model can be analyzed rigorously in terms of physical observables \cite{APS}. Although the symmetry reduction is performed at classical level as in standard mini-superspace quantization, the techniques used in LQC follow closely those of loop quantum gravity. Consequently, it can be shown that LQC quantization is inequivalent to the standard Wheeler-deWitt quantization, and thus can lead to significantly different predictions \cite{ABL}. 

The current formulation of LQC mainly deals with the quantization of {\em homogeneous} spaces but efforts are being made to incorporate inhomogeneities at the quantum level \cite{bo609} as well as within an effective classical framework that ensures an anomaly-free constraint algebra even in the presence of quantum gravity corrections \cite{Ano}. 

In this paper we review the construction of kinematical operators and the Hamiltonian constraint on a spatially-flat, homogeneous and isotropic background. The material of Sec.~\ref{vhfrw} is a self-contained review of the
literature, but also implements, with the inhomogeneous case in mind, the $\bu\sim p^{-n}$ quantization \cite{bo609} both in the gravitational and matter sectors. In Sec.~\ref{tens} we use recent findings on cosmological perturbations to compute physically relevant quantities of the early universe. In particular, we concentrate on tensor perturbations originated at an early stage in the cosmic evolution in the presence of inverse-volume corrections (holonomy corrections ignored). The near-Planckian regime of LQC typically leads to a strongly blue-tilted tensor spectrum, which is disfavoured by observations if the tensor-to-scalar ratio is not too small. However, in this regime the dynamics is intrinsically nonperturbative, as it takes place near the bounce, and the use of cosmological perturbation theory may not be completely reliable. There are also a number of other assumptions, made clear during the discussion, which may be relaxed in a more complete (and complicated) analysis. On the other hand, the quasi-classical regime of LQC is close enough to general-relativity results to be in agreement with data. In particular, the tensor index receives a small negative quantum correction, slightly enhancing the standard red tilt in slow-roll inflation. Both regimes are concrete realizations of effective dynamics with modified dispersion relations, which have been long argued to play a role in the trans-Planckian problem for inflation \cite{Bra99,BM1,MB2}.


\section{Inverse-volume and holonomy corrections}\label{vhfrw}


\subsection{Classical canonical variables, volume, and gravitational Hamiltonian}

In Ashtekar--Barbero formulation of gravity, the canonically conjugate variables are the connection $A_a^i$ and the densitized triad $E_i^a$ \cite{Ash87,Bar94}, where $a,i=1,2,3$. Here $a$ is the spatial index whereas $i$ denotes the directions in the internal space. Being conjugate variables, $A_a^i$ and $E_i^a$ satisfy the following Poisson bracket
\be\label{commr}
\{A_a^i(\mathbf{x}),\,E_j^b(\mathbf{y})\}=8\pi G\g
\delta_a^b\delta^i_j\delta(\mathbf{x},\mathbf{y})\,,
\ee
where $G$ is Newton's constant of gravitation. The physical volume can be expressed as
\be\label{vol}
\cV=\int d^3\mathbf{x}\sqrt{\det E},
\ee
where $\det E=\epsilon_{abc}\epsilon^{ijk} E_i^aE_j^bE_k^c/3!$. $\epsilon_{abc}$ is the Levi--Civita symbol:
$\epsilon_{123}=\epsilon_{231}=\epsilon_{312}=+1$, antisymmetric in pairs of indices. Repeated upper-lower indices are conventionally summed over.

The spatially-flat Friedmann--Robertson--Walker metric is $ds^2 = -dt^2 + a(t)^2dx^2$. A physical distance $d$ is determined by the dimensionless scale factor $a$ and the coordinate (comoving) length $x_0$, $d=a x_0$. The volume of a spatially-flat universe is infinite and thus, to render the volume integral \Eq{vol} well defined, we need to consider a patch of the universe with finite `fiducial' coordinate volume, say $\cV_0$. The physical volume of the patch then reads $\cV=a^3\cV_0$ \cite{ABL}.

In this background, the symmetric Ashtekar connection is expressed as $A_a^i=c\cV_0^{-1/3}\omega_a^i$, where $c\cV_0^{-1/3}=\g\dot a$ classically, $\g$ is the Barbero--Immirzi parameter \cite{Bar94,Im96}, and a dot is derivative with respect to synchronous time $t$. Here $\omega_a^i$ are the components of invariant 1-forms that satisfy $\omega_a^i\omega_b^j\delta_{ij}=\delta_{ab}$. The extrinsic curvature is $K_a^i=(A_a^i -\Gamma_a^i)/\g$.  Here $\Gamma_a^i$ is the spin connection, which vanishes for the spatially flat FRW spacetime. The symmetry-reduced conjugate momentum is $E_i^a=p\cV_0^{-2/3}e_i^a$, where $p\cV_0^{-2/3}=a^2$ (the sign of $p$ is chosen to be positive by convention) and $e_i^a$ (often denoted ${}^oe_i^a$) is the invariant vector field dual to $\omega_a^i$: $e^a_i\omega_b^i=\delta_b^a$. 

In the quantum theory the connection itself cannot be promoted as a well-defined operator; only the holonomy of the connection can be consistently quantized. The holonomy along an oriented path $e$ is an element of the $SU(2)$ group defined as 
\be
h_e \equiv {\cal P}\exp\left[\int_e d\lambda\, e^a(\lambda)
A_a^k(\lambda)\tau_k\right]\,,\label{h2}
\ee
where $d x^a/d\lambda=e^a(\lambda)$ is the tangent vector along the path parametrized by $\lambda$ and $\tau_i$, $i=1,2,3$, are the three generators of the $su(2)$ algebra in irreducible $j$-representation (of dimension $2j+1$).

For technical convenience, in the quantization of the scalar constraint operator we fix the representation to be the fundamental one, {\em i.e.}, $j=1/2$ (this choice may be justified also by theoretical considerations \cite{per05,van05}, which however are not compelling and can be bypassed). In this representation, the generators are $\tau_k=\sigma_k/(2i)$, $\sigma_k$ being Pauli matrices, and satisfy
\be\label{tauidentity}
\tau_i\tau_k=\frac{\epsilon_{ik}^{\,\,\,\,n}\tau_n}2-{\mathbb I}\frac{\delta_{ik}}4,
\ee
where ${\mathbb I}$ is the $2\times 2$ identity matrix. In the FRW case, the coordinate space is flat and the tangent
vector is the same at every point along the edge $e$. In particular, the holonomy of oriented length $\ell_0$ along the $i$-th direction is
\be\label{holonomy}
h_i^{(\mu)}=\exp(\ell_0\cV_0^{-1/3}c \tau_i)\equiv\exp(\mu c \tau_i)\,,
\ee
where the path ordering ${\cal P}$ becomes trivial and $\mu\equiv \ell_0/\cV_0^{1/3}$ is the ratio between the holonomy length and the size of the coordinate volume. To keep notation light, from now on we omit the superscript $(\mu)$ in the holonomy symbol.

In view of implementing lattice refining quantization, it is convenient to define a new pair of variables as \cite{APS,ACS}
\be
b\equiv \frac{\bu c}2\,,\qquad w\equiv \frac{2p}{\bu}\,,
\ee
where $\bu=\bu(p)$ is an arbitrary dimensionless function of $p$. Later arguments will suggest
\be\label{bu2}
\bu=\left(\frac{\Delta}{p}\right)^n\propto \left(\frac{\lp}{a}\right)^{2n},\qquad 0< n \leq 1/2\,,
\ee
where $\Delta$ is a constant of dimension $({\rm length})^2$ which can be coordinate dependent, and $\lp^2\equiv G \hbar$ is the squared Planck length. Then it is easy to show that $b$ and $w$ are also canonically conjugate and
\be
\{b,\,w\}= \frac{8\pi G \g}{3}(1+n)\,.
\ee
Defining $\lambda\equiv \mu/\bu$ as the path length in units of $\bu$, we can express the holonomy \Eq{holonomy} in terms of the new variables as
\be\label{ide}
h_i=\exp\left(2\lambda b\tau_i\right)=\mathbb{I}\cos \lambda b+2\tau_i\,\sin
\lambda b\,,
\ee
where we employed the identity Eq.~\Eq{tauidentity}. In quantizing inverse
powers of the scale factor, one faces the problem that naive inverse operators $(\hat p)^{-s}$, $s>0$, fail to remain well-defined. As a standard procedure, one rewrites inverse powers of the densitized triad using classical Poisson brackets involving positive powers of $p$. In particular, the following classical relation is useful:
\be\label{cr}
\{b,\,w^l\}=\frac{8\pi G\g}{3}(1+n)l w^{l-1},
\ee
valid for arbitrary values of $l$. Via Eqs.~\Eq{ide} and \Eq{cr}, one can express powers of $w$ as
\ba
w^{l-1} &=& \frac{\sum_i{\rm tr}(\tau_i h_i\{h_i^{-1},w^{l}\})}{8\pi G\g (1+n)l\lambda}\label{pl}\\
&=& \frac{3}{8\pi G\g l(1+n)\lambda}[\cos \lambda b\, \{\sin \lambda b,w^l\}\nonumber\\
&&\qquad-\sin \lambda b\, \{\cos \lambda b,w^l\}]\,,\label{pl2}
\ea
where we wrote explicit sums when summation convention is not clear. To get this result, we used the algebraic relation $\sum_i\tau_i\tau_i=-C_2(j){\mathbb I}$ and took the trace in the $j=1/2$ representation (for which the quadratic Casimir invariant $C_2=3/4$).

For the purpose of quantization, the Hamiltonian constraint needs to be expressed in terms of holonomies. We review this calculation \cite{boj3,boj7,Th96,Thi96}, starting from the expression of the classical Einstein--Hilbert Hamiltonian constraint
$16\pi G H_g\equiv\int d^3\mathbf{x}NC_g$:
\ba
C_g &=&\frac{E_j^aE_k^b}{\sqrt{\det E}}\left[\epsilon_i^{\,jk}F_{ab}^i-2(1+\g^2)K_{[a}^jK_{b]}^k\right]\nonumber\\
    &=&-\frac1{\g^2}\frac{E_j^aE_k^b}{\sqrt{\det E}}\,\epsilon_i^{\,jk}F_{ab}^i\label{ehh}\\
    &=& -\frac{6}{\g^2\cV_0}c^2\sqrt{p}\,,\label{li}
\ea
where $N$ is the lapse function (the choice $N=1$ corresponds to synchronous time $t$, whereas $N=a$ to conformal time $\tau$), $F_{ab}^i=\p_a A^i_b-\p_bA^i_a+\epsilon^i_{\,jk} A_a^j A_b^k$ is the gravitational field
strength, and indices in square brackets are antisymmetrized. In the second and third equalities we have made use of maximal symmetry of flat FRW backgrounds. 

To write the field strength $F_{ab}^i$ in terms of holonomies, one considers the holonomy $h_{\Box_{jk}}\equiv h_jh_kh_j^{-1}h_k^{-1}$ on a closed oriented square path whose edges have length $\mu$ and are labelled by spin indices $j,k,j,k$. Then \cite{RS1,RS2,APS}
\be
F_{ab}^i=-2\omega^j_a\omega_b^k\lim_{\ell_0\to 0} \frac{{\rm tr}(\tau_ih_{\Box_{jk}})}{\ell_0^2}\,.
\ee
Using Eq.~\Eq{ide} and $\tau_i\tau_j\tau_i=\tau_j/4$, one can show that ${\rm
tr}(\tau_{i} h_{\Box_{jk}})=-\tfrac12\epsilon_{ijk}\sin^22\lambda b$.

The terms in Eq.~\Eq{ehh} involving the densitized triad $E^a_i$ can be manipulated as follows. Starting from Eq.~\Eq{vol} and the property $\epsilon_{ibc}\epsilon^{ijk}=2\delta_{[b}^j\delta_{c]}^k$, one can prove the following formula due to Thiemann:
\be
\epsilon^{ijk}\frac{E_{j}^{a}E_{k}^{b}}{\sqrt{\det
E}}=2\epsilon^{abc}\frac{\delta\cV}{\delta
E^c_i}=\epsilon^{abc}\frac{\sum_k\omega_c^k{\rm tr} (\tau_i
h_k\{h^{-1}_k,\,\cV\})}{2\pi G\g\ell_0}\,,\label{r1}
\ee
where in the second equality we used the expression $h_k\delta h_k^{-1}/\delta A_a^i=-\ell_0 e^a_k \tau_i$.
In terms of the canonical pair $(b,w)$,
\ba
\epsilon^{ijk}\frac{E_{j}^{a}E_{k}^{b}}{\sqrt{\det
E}}&=&\epsilon^{abc}\left(\frac{\Delta^n}{2}\right)^{\frac{3}{2(1+n)}}\frac{
\omega_c^i}{2\pi G\g\ell_0}\nonumber\\
&&\times\left[\cos \lambda b\, \{\sin \lambda
b,w^{\frac{3}{2(1+n)}}\}\right.\nonumber\\
&&\qquad-\left.\sin \lambda b\, \{\cos \lambda
b,w^{\frac{3}{2(1+n)}}\}\right],\label{sqp}
\ea
in agreement with Eq.~\Eq{pl2}. The total Hamiltonian constraint \Eq{ehh} reads
\ba
C_g &=& \lim_{\ell_0\to 0}\frac{1}{\pi G\g^3\ell_0^3}\sum_{i,k'} {\rm tr}(\tau_{i} h_{\Box_{jk}})
\epsilon^{jkk'}{\rm tr} (\tau_{i} h_{k'}\{h^{-1}_{k'},\,\cV\})\nonumber\\
&=&
-\lim_{\ell_0\to 0}\frac{1}{\pi G\g^3\ell_0^3} \sin^22\lambda b\sum_i
{\rm tr} (\tau_{i} h_{i}\{h^{-1}_{i},\,\cV\})\nonumber\\
&=&\lim_{\ell_0\to 0}\frac{3}{\pi G\g^3\ell_0^3}\left(\frac{\Delta^n}{2}\right)^{\frac{3}{2(1+n)}}\sin^22\lambda b\nonumber\\
&&\qquad\times\left[\sin \lambda b\, \{\cos \lambda b,w^{\frac{3}{2(1+n)}}\}\right.\nonumber\\
&&\qquad\qquad\left.-\cos \lambda b\, \{\sin \lambda b,w^{\frac{3}{2(1+n)}}\}\right].\label{cg}
\ea


\subsection{Quantization and lattice refinement}\label{qua}

In the quantization procedure of loop quantum cosmology, the classical Poisson bracket is replaced by the commutator bracket as in standard quantum mechanics,
\be\label{quan}
\{\cdot,\cdot\}\to -\frac{i}\hbar[\cdot,\cdot]\,.
\ee
As mentioned earlier, the connection-dependent operator $\hat b$ does not have a well-defined action on the kinematical Hilbert space. Nonetheless, the other elementary variable $w$ can be promoted to a self-adjoint operator,
\be\label{wop}
w\to \hat w\equiv \frac{8\pi\lp^2\g}{3i}(1+n)\frac{d}{db}\,.
\ee
It is easy to check that the functionals $\psi^{(\lambda)}(b)=e^{i \lambda b}\equiv \langle b|\lambda\rangle$ are the eigenfunctions of $\hat w$. In Dirac notation, we can write down the action of Eq.~\Eq{wop} on the states $|\lambda\rangle$ as  
\be
\hat w |\lambda\rangle =\frac{8\pi\lp^2\g}{3}(1+n)\lambda |\lambda\rangle\equiv w|\lambda\rangle\,.
\ee
These eigenstates form a basis $\{|\lambda\rangle\}$ of the kinematical Hilbert space. On this basis, the action of a holonomy operator of edge $\lambda'$ can be constructed from the basic operator $\widehat{e^{i\lambda' b}}$, which acts simply as a translation:
\be
\widehat{e^{i\lambda' b}}|\lambda\rangle=|\lambda+\lambda'\rangle\,.
\ee
We are interested in computing the spectrum of well-defined inverse-volume operators, so we consider the values of $l$ being $0<l<1$ in Eq.~\Eq{pl2}. The ambiguity parameter $l$ determines the initial slope of the effective geometrical density. To preserve coordinate invariance when quantizing geometrical densities before symmetry reduction, $l$ must be discrete, $l_k=1-(2k)^{-1}$, $k\in \mathbb{N}$ \cite{th97,boj12}. Hence one can select the bound $1/2\leq l <1$, which is also favoured phenomenologically \cite{boj12}; a natural choice is $l=3/4$ \cite{th97,boj12}.

To quantize Eq.~\Eq{cg} in terms of elementary operators, we fix the length of the holonomy to be unity ($\lambda=1$), and consider their symmetric ordering so that Eq.~\Eq{pl2} becomes the self-adjoint operator
\be
\widehat{|w|^{l-1}} = \frac{3i}{8\pi\lp^2\g l(1+n)}\left[\widehat{\cos b}\, \widehat{|w|^l} \widehat{\sin b}-\widehat{\sin b}\,\widehat{|w|^l} \widehat{\cos b}\right].\label{plq}
\ee
The absolute value of $\hat w$ is taken in order for the eigenvalues of $\widehat{|w|^{l-1}}$ to be real. In fact, it is easy to check that the basis states $|\lambda\rangle$ are also the eigenstates of the operator Eq.~\Eq{plq}, 
\ba
\widehat{|w|^{l-1}}|\lambda\rangle &=& \frac{1}{2l} \left[\frac{8\pi\lp^2\g (1+n)}{3}\right]^{l-1}\nonumber\\
&&\qquad\times\left(|\lambda+1|^l-|\lambda-1|^l\right)|\lambda\rangle\,.\label{ww}
\ea

Although the discussion so far has been at the kinematical level, the motivations for taking $\lambda=1$ and $\bu\propto p^{-n}$ are mainly dynamical. They will be the subject of the remainder of this Subsection.

If translation invariance is broken, e.g. by inhomogeneous perturbations, the comoving volume of the system under inspection can be discretized as a lattice whose ${\cal N}$ cells have characteristic comoving size $\ell_0$ and correspond to the vertices of the spin network associated to $\cV_0$. We identify the ratio of the cell-to-lattice size $\ell_0\equiv(\cV_0/{\cal N})^{1/3}$ with the previously ad-hoc function $\bu(p)$, under the requirement that the lattice be refined in time. Hence $\bu={\cal N}^{-1/3}$.

The rescaling of the fiducial volume sometimes led to the concern that LQC quantum corrections (as well as all statements regarding the scale at which transition from semiclassical to quantum regimes occur) break conformal
invariance of a flat FRW background; scaling invariance can be realized if proper $\cV_0$ factors are considered in the observables but then the theory and its observables would depend on the choice of the fiducial volume.
However, in the presence of inhomogeneities there is no conformal freedom and, on the other hand, fluxes through two-surfaces $F=\int_\Sigma d^2\sigma E_a^i n^a \tau_i$ ($n_a$ being the normal to a surface $\Sigma$), that is, elementary areas, are determined by the inhomogeneous spin-network quantum state of the full theory associated to a given patch \cite{boj11}. This implies that to change the fiducial volume $a^3\cV_0$ would change the number of vertices of the underlying physical state. Therefore there is no scaling ambiguity in the equations of motion \cite{boj11,BH2}, although the physical observables (through $\lambda$, see below) \emph{will depend on the choice of spin-network state}. 

A feature of loop quantum gravity, which is not an assumption but a consequence of the full theory, is that the spectrum of the area operator is bounded from below by the Planck scale. Although the area spectrum of LQC does not share this property, one may draw inspiration from the full theory and consider lattice cells to be such that their kinematical area is bounded as
\be
(a \ell_0)^2\geq \Delta_{\rm Pl}\equiv 2\sqrt{3}\pi\g\lp^2\,,
\ee
where the numerical constant stems from the direct calculation in LQG \cite{ALReview,AL1}. Big-bang nucleosynthesis can place a bound over the smallest physical area, as shown in Ref. \cite{BDS}. If the inequality is
saturated (smallest possible holonomy path), the comoving cell area
\be\label{bu}
\bu^2= \frac{\Delta_{\rm Pl}}{p}\propto \left(\frac{\lp}{a\cV_0^{1/3}}\right)^2
\ee
is also the comoving area gap, that is, the smallest nonvanishing eigenvalue of the area operator measuring comoving surfaces \cite{APS,APSV,ACS}. As the Universe expands the comoving area gap shrinks to zero and the geometry is better and better described by classical general relativity, while near the big bang quantum effects become important. However, there may be more general forms than Eq.~\Eq{bu}, as the spin-network state described by the lattice can be (and usually is) excited by the action of the Hamiltonian operator on the spin vertices, increasing their number and changing their edge labels \cite{RoS,thi96}. This process has not yet been established univocally in the full theory, so it is convenient to parametrize the number of vertices as \cite{bo609} 
\be\label{Nf}
{\cal N}(t)= f_\psi(t) \cV/\lp^3,
\ee
where the (dimensionless) function $f_\psi(t)$ is state dependent (the density of vertices per Planck volume) and, \emph{by assumption}, coordinate independent; its time-dependence is inherited from the state itself. As the kinematical Hilbert space is usually factorized into gravitational and matter sectors, the problem here emerges of how to define a natural clock when matter does not enter in the definition of a (purely geometrical) spin network. This issue will require a much deeper understanding of the theory; so, as unsatisfactory as Eq.~\Eq{Nf} may be, we take it as a phenomenological ingredient in the present formulation of inhomogeneous LQC.

From the definition of $\bu$, one gets Eq.~\Eq{bu2} if $f_\psi(t)$ scales as $a^{3(2n-1)}(t)$. $\Delta$ is some physical squared length determined by the theory. Even in the improved quantization scheme, it may differ from the mass gap $\Delta_{\rm Pl}$ by some $O(1)$ constant and, as there is not yet a general consensus on this issue, we shall set $\Delta=\Delta_{\rm Pl}$ only in Eqs.~\Eq{ci}--\Eq{nat}.

Homogeneous models adopting Eq.~\Eq{bu} feature holonomies which depend on triad variables; in other words, curvature components are constrained by the area operator although this does not appear in the full constraint. On the other hand, in inhomogeneous models the dependence of the parameter $\bu$ on $p$ is implemented at state (rather than operatorial) level, in closer conformity with the full theory \cite{bo609}. For these reasons, we will keep $n$ general within a reasonable interval which can be argue to be $0<n\leq 1/2$. In fact, if $\bu=1$ ($n=0$) the states $|\lambda\rangle=|\mu\rangle$ coincide with the basis eigenstates of the momentum operator $\hat p$, with eigenvalues $w\propto p$ \cite{boj7,ABL}. The lower bound $n=0$ comes from the definition of $\bu$ (the number of vertices must grow with the physical volume) and corresponds to a configuration where the total number of vertices is kept constant during the evolution of the patch. This case leads to severe restrictions of the matter sector if the wavefunctions solving the Hamiltonian constraint are required to be normalizable and to reproduce the classical limit at large scales \cite{NS}; so it will be excluded. In the `improved quantization scheme' \cite{APS} $n=1/2$, $\{|\lambda\rangle\}$ become the eigenstate basis of the volume operator, $w\propto p^{3/2}=\cV$. In this case, the number of vertices grows linearly with the volume (constant fluxes, constant spin labels along lattice edges). There can be configurations with $n>1/2$ but they would correspond to states with high and decreasing spin labels, which are not realized generically.

To calculate the curvature at the lattice sites within $\cV_0$, we need to specify closed holonomy paths around such points. A generic holonomy is given by the composition of elementary holonomies over individual plaquettes. Therefore we set the length of the elementary holonomy to be that of the characteristic lattice cell, $\mu=\bu$.

In this case, the quantum Hamiltonian (self-adjoint) operator corresponding to Eq.~\Eq{cg} is well-defined:
\be\label{cig}
\cV_0\hat C_g = -4\,\widehat{\sin2b}\,\hat A\, \widehat{\sin2b}\,,
\ee
where
\ba
\hat A&\equiv& 
\frac{3i}{32\pi \lp^2\g^3}\left(\frac{\Delta^n}{2}\right)^{-\frac{3}{2(1+n)}}\widehat{|w|^{\frac{3n}{1+n}}}\nonumber\\
&&\qquad\times\left[\widehat{\cos b}\, \widehat{|w|^{\frac{3}{2(1+n)}}}\widehat{\sin b}\right.\nonumber\\
&&\qquad\qquad-\left.\widehat{\sin b} \,\widehat{|w|^{\frac{3}{2(1+n)}}}\widehat{\cos b}\right],
\ea
and we used the relations
\be
p=\left(\frac{\Delta^n}{2}w\right)^{\frac{1}{1+n}}\,,\qquad \bu=2\left(\frac{\Delta^n}{2w^n}\right)^{\frac{1}{1+n}}.
\ee
Acting on a state $|\lambda\rangle$, which is an eigenstate of $\hat A$ with eigenvalue $A_\lambda$, one has
\be
\cV_0\hat C_g|\lambda\rangle = A_{\lambda+2}|\lambda+4\rangle-(A_{\lambda+2}+A_{\lambda-2})|\lambda\rangle+A_{\lambda-2}|\lambda-4\rangle\,,
\ee
where 
\ba
A_\lambda &=& \frac{1+n}{8\g^2}\left(\frac{\Delta^n}{2}\right)^{-\frac{3}{2(1+n)}}
\left[\frac{8\pi\lp^2\g (1+n)}{3}\right]^{\frac{1+4n}{2(1+n)}}\nonumber\\
&&\qquad\times\lambda^{\frac{3(1+2n)}{2(1+n)}}\left(\left|1+\frac1\lambda\right|^\frac{3}{2(1+n)}-\left|1-\frac1\lambda\right|^\frac{3}{2(1+n)}\right).\nonumber\\
\ea
Finally, we define the function
\ba
\a&\equiv& \frac{2\g^2}{3}\frac{\bu^2}{\sqrt{p}}A_\lambda\nonumber\\
&=&\frac{8\g^2}{3}\left(\frac{\Delta^n}{2}\right)^{\frac{3}{2(1+n)}}\left[\frac{8\pi\lp^2\g (1+n)\lambda}{3}\right]^{-\frac{1+4n}{2(1+n)}}A_\lambda\nonumber\\
&=&\frac{1+n}{3}\lambda\left(\left|1+\frac1\lambda\right|^\frac{3}{2(1+n)}-\left|1-\frac1\lambda\right|^\frac{3}{2(1+n)}\right),
\ea
which is nothing but a correction function to the eigenvalue of the operator
$\widehat{p^2/p^{3/2}}$ (quantization of Eq.~\Eq{sqp}) on $|\lambda\rangle$,
$\widehat{p^2/p^{3/2}}|\lambda\rangle=\a\sqrt{p}|\lambda\rangle$.
Note that, a priori, another legitimate choice is to write Eq.~\Eq{r1} as
\be
\sqrt{p}\propto \frac{\cV^{1-r}}{r}\frac{\delta\cV^r}{\delta p}\propto \frac{\cV^{1-r}}{r\ell_0}\sum_i{\rm tr} (\tau_i h_i\{h^{-1}_i,\,\cV^r\})\,,\label{rgen}
\ee
which introduces another ambiguity $0<r\leq1$. Then it is easy to show that
\ba\label{alpha}
\a
&=&\frac{1+n}{3r}\lambda\left(\left|1+\frac1\lambda\right|^\frac{3r}{2(1+n)}-\left|1-\frac1\lambda\right|^\frac{3r}{2(1+n)}\right).
\ea


\subsection{Effective background dynamics}\label{c}

Effective cosmological equations of motion are derived from the expression of the effective Hamiltonian constraint. To get the latter, typically one considers a semiclassical state $|\psi_{\rm sc}\rangle$ peaked around some point
$(w,b)$ in the classical phase space. In this state one then computes the expectation value of the Hamiltonian constraint operator using appropriate inner product. Accordingly, for the gravitational Hamiltonian operator
$\hat H_g$ we may write
\be\label{Hg}
\langle\psi_{\rm sc}|\hat H_g|\psi_{\rm sc}\rangle \approx -\frac{3N}{8\pi
G}\,\a\sqrt{p}\,\frac{\sin^2 \bu c}{\bu^2\g^2}\,,
\ee
where we have approximated the expectation value of the operators in Eq.~\Eq{cig} as
$\langle\psi_{\rm sc}|\widehat{\sin \bu c}\, \hat A\, \widehat{\sin \bu
c}|\psi_{\rm sc}\rangle \approx A_\lambda\sin^2 \bu c$.

Including matter components, imposition of the effective Hamiltonian constraint (that is, variation of the total effective Hamiltonian $H_{\rm tot}=\langle \hat H_g\rangle+\langle\hat H_{\rm mat}\rangle$ with respect to $N$) yields
\be\label{frw0}
\a\,\frac{\sin^2 \bu c}{(a\bu\g)^2}=\frac{8\pi G}{3}\,\rho\,,
\ee
where $\rho= p^{-3/2}\delta \langle \hat H_{\rm mat}\rangle/\delta N$. When $\a=1$ and the matter sector contains only a massless free scalar field (potential $V(\varphi)=0$), then Eq.~\Eq{frw0} is exact \cite{bo10}. In general, however, the evolution of a finitely-spread semiclassical state will produce quantum fluctuations leading to additional corrections to Eq.~\Eq{frw0}. Here, for simplicity, we shall ignore such contributions; these were discussed in Refs. \cite{BHS,bo11}. The Hamilton equation of motion for the densitized triad $\dot p=\{p,H_{\rm tot}\}$
gives the Hubble parameter
\be\label{hub}
H\equiv \frac{\dot a}{a} = \a \,\frac{\sin 2\bu c}{2a\bu\g}\,.
\ee
In the classical limit, $c\to\g\dot a$. Combining Eqs.~\Eq{frw0} and \Eq{hub} one gets the Friedmann equation
\be\label{frw}
H^2=\frac{8\pi G}{3}\,\rho\left(\a-\frac{\rho}{\rho_{\rm c}}\right),
\ee
where
\be\label{rho*}
\rho_{\rm c} \equiv \frac{3}{8\pi G\g^2 \bu^2 p}\,.
\ee
From the definition of the number of cells/vertices,
\be
G\rho_{\rm c} \propto \left(\frac{\cN}{\cV}\right)^{2/3}=\frac{f_\psi^{2/3}(t)}{\lp^2}.
\ee
The improved quantization scheme (\cite{APS}, Eq.~\Eq{bu}) is the only one where $\rho_{\rm c}$ is constant \cite{APS,sin06,SVV}, as $\cN\propto \cV$. Otherwise, it will depend on the function $f_\psi(t)$ wherein we have encoded our ignorance about the underlying state \cite{bo11}. In both cases, the critical density is a number density which does \emph{not} depend on the size of the fiducial volume, nor on coordinates.

We assume validity of the semiclassical approximation and that the semiclassical wave packet of the universe does not spread appreciably, so we presently stick to Eq.~\Eq{frw} also in the presence of a nontrivial potential. For a scalar field $\vp$ with momentum $\pi_\vp$ and potential $V$, matter Hamiltonian is
\be
H_{\rm mat}=N\left(\frac{\pi_\vp^2}{2p^{3/2}}+p^{3/2}V\right)\,,
\ee
which upon quantization can be written as
\be\label{SFHamOp}
\hat H_{\rm mat}=\hat N\left(\frac12\widehat{p^{-3/2}}\widehat{\pi_\vp^2}+\widehat{p^{3/2}}\hat{V}\right)\,.
\ee
The effective matter Hamiltonian can be viewed as the expectation value of
the matter Hamiltonian operator in the semiclassical state $|\psi_{\rm sc}\rangle$ (whereon $\hat\vp$ acts multiplicatively):
\be
\langle\psi_{\rm sc}|\hat H_{\rm mat}|\psi_{\rm sc}\rangle \approx 
N\left(\frac{\nu \pi_\vp^2}{2p^{3/2}}+p^{3/2}V\right)\,,
\ee
where we have neglected quantum fluctuations from the matter sector and $\nu \equiv \langle\lambda|\widehat{p^{3/2}}\widehat{p^{-3/2}}|\lambda\rangle=(\langle\lambda|\widehat{w^{1-l}}\widehat{w^{l-1}}|\lambda\rangle)^{3/[2(1-l)(1+n)]}$. Using Eq.~\Eq{ww},
\be
\nu=\left[\frac{\lambda}{2l}\left(\left|1+\frac1\lambda\right|^l-\left|1-\frac1\lambda\right|^l\right)\right]^{\tfrac{3}{2(1-l)(1+n)}}.
\ee
The equation of motion $\dot\vp=\{\vp,H_{\rm tot}\}$ yields $\pi_\vp=p^{3/2}\dot\vp/(N\nu)$, while $\dot\pi_\vp=\{\pi_\vp,H_{\rm tot}\}$ leads to
\be\label{kg}
\ddot\vp+3H\left(1-\frac{\dot\nu}{3H\nu}\right)\dot\vp+\nu V_{,\vp}=0\,.
\ee
As $\nu\geq 0$ has a maximum at $\lambda=1$ and then decreases down to unity for large $\lambda$, the friction term in Eq.~\Eq{kg} changes sign during the evolution of the universe, the first stage being of superacceleration, as anticipated.

Since the momentum operator is $\hat w= 2\widehat{p/\bu}$, the total $p$-dependence of $\lambda$ is
\be
\lambda =\frac{3\sqrt{3}}{2(1+n)}\frac{\Delta}{\Delta_{\rm Pl}}\left(\frac{p}{\Delta}\right)^{1+n}.
\ee
Then the eigenvalues of the volume operator $\hat\cV$ are proportional to $\lambda^{3/(2+2n)}$, and the classical limit (large-volume approximation) corresponds to $\lambda\gg 1$. Consistently, in this regime the eigenvalues of $\widehat{|w|^{l-1}}$ are $\approx w^{l-1}$. Then Eq.~\Eq{alpha} can be approximated as
\ba
\a&\approx& 1+\left[\frac{3r}{2(1+n)}-2\right]\left[\frac{3r}{2(1+n)}-1\right]\frac{1}{6\lambda^2}\nonumber
\\
&\equiv& 1+\a_c\left(\frac{\sqrt{\Delta}}{a\cV_0^{1/3}}\right)^{c},\label{ca}
\ea
where
\be
c=4(1+n)\,,
\ee
and
\be
\a_c=\frac{[3r-4(1+n)][3r-2(1+n)]}{3^42}\left(\frac{\Delta_{\rm Pl}}{\Delta}\right)^2.
\ee
Near Planck scale ($\lambda\ll1$),
\be
\a\approx \lambda^{2-\frac{3r}{2(1+n)}}\equiv \a_q\left(\frac{a\cV_0^{1/3}}{\sqrt{\Delta}}\right)^{q_\a},\label{qa}
\ee
where
\be
q_\a=4(1+n)-3r\,,\qquad\a_q=\left[\frac{3\sqrt{3}}{2(1+n)}\frac{\Delta}{\Delta_{\rm Pl}}\right]^{\frac{q_\a}{2(1+n)}} \,.
\ee
On the other hand, when $\lambda\gg 1$,
\be
\nu\approx 1+\frac{2-l}{2(1+n)\lambda^2}\equiv 1+\nu_c\left(\frac{\sqrt{\Delta}}{a\cV_0^{1/3}}\right)^{c},\label{cn}
\ee
where
\be
\nu_c=\frac{2(1+n)(2-l)}{27}\left(\frac{\Delta_{\rm Pl}}{\Delta}\right)^2.
\ee
Near Planck scale ($\lambda\ll1$),
\be
\nu\approx \lambda^{\frac{3(2-l)}{2(1-l)(1+n)}}
\equiv \nu_q\left(\frac{a\cV_0^{1/3}}{\sqrt{\Delta}}\right)^{q_\nu},\label{qn}
\ee
where
\be
q_\nu=\frac{3(2-l)}{1-l},\qquad \,\nu_q=\left(\frac{3\sqrt{3}}{2+2n}\frac{\Delta}{\Delta_{\rm Pl}}\right)^{\frac{q_\nu}{2(1+n)}}.
\ee
Assuming $\Delta =\Delta_{\rm Pl}$, the parameter ranges are
\ba
&&4<c\leq 6\,,\label{ci}\\
&&-0.01\approx -\frac1{162}<\a_c<\frac19\approx 0.11\,,\label{alc}\\
&&1.61\approx\frac{3^{3/4}}{\sqrt{2}}<\a_q< \frac{27}{4}\approx 6.75,\\
&& 1<q_\a< 6\,,\label{qan}\\
&&0.07\approx \frac2{27}<\nu_c<\frac16\approx 0.17\,,\\
&&\nu_q\geq 3^{3/2}\approx 5.20,\qquad q_\nu\geq 9\,.
\ea
Although one can resort to different quantization schemes, general arguments show that Eqs.~\Eq{ca}, \Eq{qa}, \Eq{cn}, and \Eq{qn} maintain the same structure, where the coefficients $c$, $q_\a$, and $q_\nu$ are robust in the choice of the parameters, inasmuch as their order of magnitude does not change appreciably \cite{BHKS}. All these parameters can be set to their `natural' values, which are dictated by the form of the Hamiltonian or other considerations. From the calculation leading to $\a$ and $\nu$, it is clear that the natural choice is
\be\label{natu}
l=3/4\,,\qquad r=1\,,\qquad n=1/2\,,
\ee
which gives $c=6$ and
\ba
&&\a_c=0\,,\qquad\quad\qquad \a_q=\sqrt{3}\,,\qquad\qquad\quad\quad q_\a=3\,,\nonumber\\
&&\nu_c=\frac5{36}\approx 0.14 \,,\quad \nu_q=3^{5/2}\approx 15.59\,,\qquad q_\nu=15\,.\nonumber\\\label{nat}
\ea
At this point it may be useful to summarize different parameter choices in the literature: the same sequence of steps we reviewed was followed in Refs. \cite{boj8,boj9} (arbitrary $j$, $l=1/2$, $n=0$; $\a=1$), \cite{boj11} (arbitrary $j$ and $l$, $n=0$; $\a=1$), \cite{APS} ($j=1/2$, $l=1/2$, $r=1$, $n=1/2$), \cite{BHKS} (arbitrary $j$, $l$, $r=1$, $n=1/2$), and partly \cite{NS} ($j=1/2$, arbitrary $l$, $r=1$, arbitrary $n$).

All quantum corrections depend on the ratio
\be\label{main}
\frac{\lp}{\cV^{1/3}}=\left(\frac{f_\psi}{\cN}\right)^{1/3}.
\ee
In a purely homogeneous universe, the left-hand side is volume dependent and inverse-volume corrections will feature $\cV$, even in the case $n=1/2$ (when $r=1$, $\a=1$ but in general $r\neq 1$; $\nu$ is never constant for any allowed value of the parameters). For a compact universe (e.g., a sphere or a torus) the fiducial volume is identified with the total physical volume, which is given by the theory; so no issue arises in this case. Otherwise, the fiducial volume is arbitrary and hence unphysical. However, in the lattice-refinement picture the same quantity is determined by the inhomogeneous state through the number of vertices $\cN$ and the function $f_\psi$. Inhomogeneity is not simply invoked to justify Eq.~\Eq{main} but can be implemented via a spatially local metric as in the covariant \cite{ElB,BDE,BED,LV1,LV2} and separate universe approach \cite{WMLL} to cosmological nonlinear perturbations.


\subsection{Background expansion}

In order to solve analytically the perturbed equations of motion of the next Section, it will be necessary to choose a simple background. In standard inflation, this is achieved either in de Sitter spacetime ($a=e^{Ht}$, $H$ constant) or when the scale factor has a power-law behaviour: 
\be\label{at}
a(t)\propto t^{\tilde p},\qquad {\tilde p}>0\,.
\ee
We should mention here that $a$ is singular as $t\to 0$; thus, close to bounce point Eq.~\Eq{at} fails. We comment on this issue in the discussion.

In conformal time $\tau\equiv\int dt/a$ Eq.~\Eq{at} would be $a\propto\tau^{{\tilde p}/(1-{\tilde p})}$ if
${\tilde p}\neq 1$, but for simplicity we define $p={\tilde p}/(1-{\tilde p})$
and make the ansatz (up to an arbitrary normalization costant)
\be\label{atp}
a=\tau^p.
\ee
Below, the symbol $p$ should not to be confused with the triad, which will be always written as $a^2$.
Then $\cH\equiv \p_\tau a/a =aH=p/\tau$. The first slow-roll parameter reads
\be
\epsilon=-\frac{\dot H}{H^2}=1+\frac1p\,.
\ee
Inflation occurs for $p<-1$, superinflation when $-1<p<0$, and normal expansion when $p>0$.
In particular,
\be
\tau=-\frac{1}{\cH(1-\epsilon)}=\frac{p}{\cH}\,.
\ee
In conformal time, the de Sitter solution ($p= -1$) is $a=(-H\tau)^{-1}$, $\cH=-1/\tau$. de Sitter and power-law solutions can be used as sensible backgrounds only if they are classically stable against homogeneous perturbations. This check was done in Ref. \cite{CMNS}.
 

\section{Tensor perturbations and observables}\label{tens}

Metric perturbations can be decomposed into three different modes: scalar, vector and tensor. At linearized level, these modes do not couple to each other and gauge transformations do not mix them among themselves. Thus, they can be studied independently. 

In considering quantum corrections to the effective Hamiltonian constraint, one needs to ensure that the modified constraint algebra remains anomaly-free in the canonical approach. In classical theory, closure of the constraint algebra is guaranteed by general covariance. However, in the presence of quantum corrections, one needs to explicitly show that the constraint algebra can be made anomaly-free. Such analyses have been recently performed in the presence of inhomogeneous perturbations including only inverse-volume corrections \cite{Ano}.

Due to inverse-volume and holonomy corrections, it was shown that vector perturbations are suppressed even faster than in classical cosmology \cite{BH1}. The derivation of gauge-invariant variables for the scalar sector are still under investigation. So here we focus only on tensor-mode dynamics with inverse-volume corrections. From now on we fix $\cV_0=1$; volume prefactors will be restored at the end of the calculation.

The linearized equation of motion for tensor modes has been computed \cite{BH2}. There, triad and connection components are separated into a FRW background and an inhomogeneous perturbation,
\be
E_i^a=a^2\delta_i^a+\delta E_i^a\,,\qquad A_a^i=c\delta_a^i+(\delta\Gamma_a^i+\g \delta K_a^i)\,.
\ee
One can see that
\be\label{delEK}
\delta E_i^a= -\frac12 a^2 h_i^a\,,\qquad \delta K_a^i=\frac12 \left(\frac1\a\p_\tau h_i^a+\frac{c}{\gamma}h_i^a\right)\,,
\ee
where $h_i^a$ is the transverse traceless part of the perturbed 3-metric and the curvature and triad perturbations are canonically conjugate:
\be\label{deldel}
\{\delta K_a^i({\bf x}),\delta E_j^b({\bf y})\}=8\pi G \delta_a^b\delta_j^i\delta({\bf x},{\bf y})\,.
\ee
One can go to momentum space and denote with $h_k$ either of the two polarization modes of the graviton $h_i^a$, where $k$ is the comoving wave number of the perturbation. When only inverse-volume corrections are taken into account and in the absence of anisotropic stress, the equation of motion for $h_k$ is \cite{BH2}
\be\label{mue}
\p_\tau^2 h_k+\cH\left(2-\frac{d \ln\a}{d\ln a}\right)\p_\tau h_k+\a^2k^2 h_k=0\,.
\ee
Near Planck scale
\be\label{mueq}
\p_\tau^2 h_k+\cH(2-q_\a)\p_\tau h_k+\a_q^2\left(\frac{a}{\sqrt{\Delta}}\right)^{2q_\a}k^2 h_k=0\,,
\ee
while in the quasi-classical regime
\ba
&&\p_\tau^2 h_k+\cH\left[2+c\a_c\left(\frac{\sqrt{\Delta}}{a}\right)^c\right]\p_\tau h_k\nonumber\\
&&\qquad+\left[1+2\a_c\left(\frac{\sqrt{\Delta}}{a}\right)^c\right]k^2 h_k\approx 0\,,\label{muec}
\ea
where in the last equation we have retained only leading-order terms in $\a_c$.


\subsection{Near-Planck regime}\label{nepr}

To solve Eq.~\Eq{mueq}, we make some field and variable redefinitions. First, we define the Mukhanov field $w_k\equiv a h_k$, the constant $\nu\equiv 1/2-p/(1+pq_\a)$ (where $pq_\a\neq -1$), and recast Eq.~\Eq{mueq} in the variable $z\equiv \int d\tau \a=\tau\a/(1+pq_\a)$. Then
\be\label{xx}
\p_z^2 w_k+\left(k^2-\frac{4\nu^2-1}{4z^2}\right)w_k=0\,,
\ee
and the solution reads (Ref. \cite{GR} formula 8.491.5)
\be
w_k=C_1 \sqrt{-kz}\, H^{(1)}_{\nu}(-kz)+ C_2 \sqrt{-kz}\, H^{(2)}_{\nu}(-kz)\,,
\ee
where $H^{(i)}_\nu$ are Hankel functions of the first and second kind. Some useful formul\ae\ to get Eq.~\Eq{xx} are:
\ba
&&\p_\tau h_k=\frac1a \left(\p_\tau w_k-\cH w_k\right)\,,\\
&&\p_\tau^2h_k=\frac1a \left[\p_\tau^2w_k-2\cH \p_\tau w_k+\epsilon\cH^2w_k\right]\,,\\
&&\p_\tau=\a\p_z\,,\qquad \p_\tau^2=\a^2\p_z^2+\a q_\a\cH\p_z.
\ea
In the classical limit, $z\to \tau$ and one recovers the usual result. Taking $C_2=0$ (in order to get the correct asymptotic behaviour at small scales, advancing plane wave), one can study the large- and short-wavelength limits of the solution ($\nu>0$)
\ba
w_k &\sim& -iC_1\,\frac{2^\nu\Gamma(\nu)}{\pi}\,(-kz)^{1/2-\nu}\,,\quad |kz|\ll 1\,,\\
w_k &\sim& C_1\, \sqrt{\frac{2}{\pi}}e^{-i\left(kz+\frac\pi2\nu+\frac\pi4\right)}\,\, \,,\qquad |kz|\gg 1\,.\label{kzl}
\ea
The tensor spectrum is conventionally defined as
\be
A_T^2\equiv\frac{{\cal P}_{h}}{100}\equiv \frac{k^3}{200\pi^2a^2}\sum_{+,\times}\left\langle|\hat u_{k\ll\cH}|^2\right\rangle\Big|_{k=k_*},
\ee
where $k_*$ is the wavenumber at horizon crossing, the solution is at long wavelengths, the sum is over the two polarization modes, and angular brackets denote the vacuum expectation value of the operator $\hat u_k\equiv a\hat h_k$. The constant $C_1$ is determined by choosing the Bunch--Davis vacuum (asymptotically Minkowski) in the classical regime, $w_k\sim e^{-ikz}/\sqrt{2k}$, as follows. Taking Eqs.~\Eq{delEK}, \Eq{deldel}, and \Eq{quan}, the commutation relation between $\hat u_k$ and its conformal derivative reads
\be\label{1}
\left[\hat u_{k_1},\,\p_\tau\hat u_{k_2}\right]=32\pi\lp^2 i \a\delta(k_1,k_2)\,.
\ee
The Mukhanov variable is expanded in terms of creation and annihilation operators,
\be\label{co}
\hat u_k=w_k a_k+w_k^* a_k^\dagger,
\ee
where $*$ indicates complex conjugate, $w_k$ is the classical solution, and
\ba
&&[a_{k_1},\,a_{k_2}^\dagger] = \delta(k_1,k_2)\,,\label{3}\\
&&[a_{k_1},\,a_{k_2}] = 0 =[a_{k_1}^\dagger,\,a_{k_2}^\dagger]\,.
\ea
Plugging equation \Eq{co} into \Eq{1} and using equation \Eq{3}, one gets 
\be\label{bdv}
w_k \p_\tau w_k^*-w_k^*\p_\tau w_k=i(32\pi\lp^2)\a.
\ee 
Hence, from Eqs.~\Eq{kzl} and \Eq{bdv} $|C_1|=\sqrt{8\pi^2\lp^2/k}$.

Horizon crossing is defined at the moment when perturbations are frozen, namely,
\be
k_*=\frac{\sqrt{4\nu^2-1}}{2z}=\frac{\cH}{\a}\sqrt{1-q_\a-\frac1p}\,.
\ee
In the classical limit, $k_*\to\sqrt{2} aH$, as expected. The above expression is well defined only if $p>1/(1-q_\a)$, which, according to the value of $q_\a$, will determine (compatibly with the stability of the background solution) whether the universe superaccelerates or not. In order to avoid interpretational issues \cite{MN}, the stronger condition $p>-1/q_\a$ must hold (superinflation, if the universe accelerates), so that the time variable $z$ flows along the same direction at $\tau$ and modes exit (rather than enter) the Hubble horizon at crossing. Then,
\ba
A_T^2 &=& \frac{\lp^22^{2\nu+1}\Gamma^2(\nu)}{25\pi^2}\frac{(\nu^2-1/4)^{3/2-\nu}}{(\nu-1/2)^2}\frac{H^2}{\a^2}\nonumber\\
&\propto& k^{2(1+p+pq_\a)/(1+pq_\a)},
\ea
and the tensor spectral index is
\be
n_T\equiv\left.\frac{d\ln A_T^2}{d\ln k}\right|_{k=k_*}=\frac{2(\epsilon+q_\a)}{\epsilon+q_\a-1},
\ee
In the quasi-de Sitter limit $p\approx-1$, $n_T\approx 2q_\a/(q_\a-1)$, leading to a blue-tilted spectrum $n_T>12/5$. The tensor index is greater than $1$ also when the universe is in deep superacceleration ($\epsilon\ll -q_\a$, $p\to 0^-$), $n_T\approx 2$. Prior to the calculation of tensor perturbations, some toy models already favoured a strong blue tilt of the spectrum \cite{MS1,MS2}. This state of affairs is typical of LQC inflation and superinflation near Planck regime, since an almost scale-invariant tensor spectrum would require a certain degree of fine tuning, $\epsilon\approx -q_\a$, which however could spoil scale-invariance in the scalar sector \cite{CMNS,MN,LQC1}. 

The spectrum of primordial gravitational waves is not directly observed as after inflation it evolves to a stochastic background $\Omega_{\rm gw}$ through radiation and matter eras. This process can be modelled in a transfer function $T(k)$ \cite{TWL}, so that \cite{CE}
\be
\Omega_{\rm gw}=\frac{1}{\rho_0}\frac{d\rho_{\rm gw}}{d\ln k}= T^2(k)A_T^2\,,
\ee
where $\rho_0=3H^2_0/(8\pi G)$ is the critical energy density of the universe ($H_0$ is today's Hubble parameter), and $\rho_{\rm gw}$ is the energy density of the gravitational waves. Then \cite{SB}
\be\label{ff}
n_T\approx \frac{1}{\ln f-\ln f_0}\ln\left(2.29\times 10^{14}\frac{h^2\Omega_{\rm gw}}{r}\right)\,,
\ee
where $f= k/(2\pi)$ is the frequency of the signal, $f_0=a_0H_0/(2\pi)=3.10\times 10^{-18}$ Hz, $h=0.716$, and $r\equiv A_T^2/A_S^2$ is the ratio between tensor and scalar ($A_S^2$) amplitudes. Observations of pulsar timing, interferometer experiments (LIGO, LISA), and the theory of big-bang nucleosynthesis (BBN) can place strong constraints on the tensor index \cite{SB}. For instance, taking the upper bound $r<0.30$, from pulsar timing $n_T\lesssim 0.79$, while from BBN $n_T\lesssim 0.15$. These values are incompatible with the above predictions of near-Planckian LQC. Therefore, modulo some important issues we shall comment later, we might conclude that a near-Planckian accelerating phase might have occurred only at very early times (unobservably large scales), and for a short period.

In homogeneous models the duration of this regime was held to depend on the spin representation of the holonomies, small $j$ implying a very short superinflationary period \cite{boj12}. In inhomogeneous situations, this problem is reinterpreted and relaxed in terms of the lattice embedding of Ref. \cite{bo609}. The volume spectrum depends on the quadratic Casimir in $j$ representation: $ \bu^{-n}\sim \cV^{2/3}\sim \sqrt{C_2(j)}\sim j$. A higher-$j$ effect can be obtained as a refinement \cite{BCK} of the lattice (smaller $\bu$), thus allowing for long enough superacceleration. A change in $\bu(p)$ can be achieved by varying the comoving volume $\cV_0$. This is an arbitrary operation in pure FRW, while in inhomogeneous models $\bu$ is a physical quantity related to the number of vertices of the underlying reduced spin-network state. As long as a calculation of this effect from the full theory is lacking, we will not be able to predict the duration of the small-volume regime.


\subsection{Quasi-classical regime}

If one assumes that inflation takes place far enough from the Planck era, one can look for solutions which are perturbative in the small parameter $\a_c$, $w_k=\sum_n\a_c^n w^{(n)}_k$. The natural choice Eq.~\Eq{nat} coincides with classical gravity and obviously it can fit experimental data; so we shall concentrate on the case $\a_c\neq 0$. Rewriting Eq.~\Eq{muec} as
\ba
&&\p_\tau^2w_k+c\cH(\a-1)\p_\tau w_k+\{(2\a-1)k^2\nonumber\\
&&\qquad\qquad+\cH^2[\epsilon-2-c(\a-1)]\}w_k=0\,,
\ea
it is sufficient to consider the ansatz
\be
w_k=w_k^{(0)}+\a_c w_k^{(1)}\,,
\ee
and solve at zero and lowest order in $\a_c$. One obtains two equations,
\ba
&& \p_\tau^2 w_k^{(0)}+\left[k^2+\cH^2(\epsilon-2)\right]w_k^{(0)}=0\,,\label{w0}\\
&& \p_\tau^2 w_k^{(1)}+\left[k^2+\cH^2(\epsilon-2)\right]w_k^{(1)}+r(\tau)=0\,,\label{w1}\\
&& r(\tau)\equiv \left(\frac{\sqrt{\Delta}}{a}\right)^c\left[c\cH\p_\tau w_k^{(0)}+(2k^2-c\cH^2)w_k^{(0)}\right]\,,\nonumber\\\label{erre}
\ea
the first being the usual Mukhanov equation in general relativity and the second having a source term. These equations can be solved simultaneously and exactly (even at higher orders in $\a_c$, using Eq.~\Eq{mue}) with standard techniques for linear differential equations with variable coefficients; however, the result for $w^{(1)}_k$ would be uninstructively complicated and we do not show it here. Rather, we consider the large-scale ($k\ll {\cal H}$) and small-scale ($k\gg {\cal H}$) regimes separately, and fix normalization constants by joining asymptotic solutions at horizon crossing. Equations \Eq{w0} and \Eq{w1} make it happen when
\be\label{hcros}
k_*=\cH\sqrt{1-\frac1p}\,,
\ee
as in the classical case. In the large-scale limit the solution is very simple and coincides with the standard constant mode
\be\label{wls}
w_{k\ll \cH}=C_1(1+\a_cC_2)\tau^p\,,
\ee
where we have ignored $O(\tau^{1-p})$ decaying terms (we assume inflationary dynamics, $p<-1$). From Eq.~\Eq{bdv}, at lowest order the usual relation
$w_k^{(0)} \p_\tau w_k^{(0)*}-w_k^{(0)*}\p_\tau w_k^{(0)}=i(32\pi\lp^2)$ yields at small scales
\be\label{w0sol}
w_{k\gg\cH}^{(0)}= \sqrt{\frac{16\pi\lp^2}{k}}\,e^{-ik\tau}\,.
\ee
The solution of Eq.~\Eq{w1} is given by the general solution of the homogeneous equation plus a particular solution of the inhomogeneous one. In the large $k$ limit ($\cH^2$ terms neglected in Eqs.~\Eq{w1} and \Eq{erre}), one can show that
\be\label{w1sol}
w_k^{(1)}\approx C_3e^{-ik\tau}+C_4e^{ik\tau}+w_k^{(0)}\frac{ik\tau}{cp-1}\left(\frac{\sqrt{\Delta}}{a}\right)^c\,.
\ee
The linear (in $\a_c$) part of Eq.~\Eq{bdv} is
\ba\label{wro}
&&w_k^{(0)} \p_\tau w_k^{(1)*}-w_k^{(0)*}\p_\tau w_k^{(1)}+w_k^{(1)} \p_\tau w_k^{(0)*}\nonumber\\
&&\qquad-w_k^{(1)*}\p_\tau w_k^{(0)}=i(32\pi\lp^2)\left(\frac{\sqrt{\Delta}}{a}\right)^c.
\ea 
Plugging Eqs.~\Eq{w0sol} and \Eq{w1sol} in Eq.~\Eq{wro}, one can fix the coefficients $C_3=0=C_4$. Therefore the solution at small scales is
\be\label{wtot}
w_{k\gg\cH}=w_{k\gg\cH}^{(0)}\left[1+\a_c\frac{ik\tau}{cp-1}\left(\frac{\sqrt{\Delta}}{a}\right)^c\right],
\ee
while the solution at large scales is Eq.~\Eq{wls} with
\ba
C_1(k) &=&\sqrt{\frac{16\pi\lp^2}k}\,\frac{e^{-ik\tau_*}}{a_*}\equiv \tilde C_1 k^{p-1/2}\,,\\
C_2(k) &=&\frac{ik_*\tau_*}{cp-1}\left(\frac{\sqrt{\Delta}}{a_*}\right)^c\equiv \tilde C_2 k^{cp}\,,
\ea
where $\tau_*=\sqrt{p(p-1)}/k$ and $a_*$ are defined through Eq.~\Eq{hcros}. As expected, the LQC correction term in Eq.~\Eq{wtot} decays (as $\tau^{1-cp}$) and becomes asymptotically negligible with respect to the purely classical mode.

The tensor amplitude is
\be
A_T^2=\frac{4\lp^2}{25\pi}\frac{k^{2(1+p)}}{[p(p-1)]^p}(1+\delta_{\rm Pl})\,,
\ee
where
\be\label{deltalp}
\delta_{\rm Pl}\equiv \a_c^2|\tilde C_2|^2 k^{2cp}\,.
\ee
The tensor index reads, for small $\delta_{\rm Pl}=O(\lp^{2c})$,
\be
n_T\approx 2(1+p+cp\delta_{\rm Pl})=\frac{-2(\epsilon+c\delta_{\rm Pl})}{1-\epsilon}\,.
\ee
Up to an $O(1)$ factor,
\be\label{dell}
\delta_{\rm Pl}\propto \frac{\a_c^2}{(1-cp)^2} \left[\frac{\lp}{\cV_*^{1/3}}\right]^{2c}\,,
\ee
which can be recast in terms of $f_\psi$ and $\cN$ as in Eq.~\Eq{main}. Because the present status of the theory is not advanced enough to know the details of the semiclassical state, inhomogeneous loop quantum cosmology does not yet have enough predictive power to make a unique statement about the magnitude of $\delta_{\rm Pl}$ for a given pivot scale $k_*$. However, Eqs.~\Eq{main} and \Eq{dell} illustrate the qualitative behaviour of quantum corrections in the large class of semiclassical states modelized by lattice refinement. Whether and how these states emerge in the full theory is a problem which goes beyond the scope of this paper. 

Actually, there is a natural scale that may fix $\cV^{1/3}$ on any time slice: namely, the particle horizon. For a large universe this is approximated by the Hubble radius $H^{-1}$ (we have implicitly assumed throughout the paper that the fiducial volume is much larger than the Hubble volume). In Ref. \cite{BG}, the effect of holonomy corrections were considered in tensor cosmological observables. In particular, the inflationary tensor index gets an extra contribution $\delta_{\rm hol}\propto(\lp H_*)^2$. Then $\delta_{\rm Pl}\sim (\lp H_*)^{2c}\ll \delta_{\rm hol}$ and the scenario is dominated by holonomy corrections. Taking the grand-unification scale $H_*\sim 10^{14} \div 10^{17}~{\rm GeV}$, $\delta_{\rm hol}\sim 10^{-10}\div 10^{-4}$, so none of the corrections is observable. The same conclusion is reached even when taking the largest possible inverse-volume correction. By definition of quasi-classical regime, $\cV_*^{1/3}>\lp$. From Eqs.~\Eq{ci} and \Eq{alc}, one has the upper bound $\delta_{\rm Pl}\lesssim O(10^{-3})$.   


\section{Conclusions}

In summary, we showed that the near-Planckian regime of LQC with inverse-volume corrections typically (but not always) leads to a strongly blue-tilted tensor spectrum, which is disfavoured by observations. Separate studies on the
holonomy-corrected tensor spectrum \cite{BG} are not too encouraging either, as the tensor index is blue-tilted even in that case \cite{BG,MS3} (during a slow-roll phase, $n_T$ is nearly scale invariant, therefore the inverse-volume contribution always dominates). Thus, the LQC superinflationary phase may not be a phenomenologically viable scenario to explain the observational bounds on the tensor index. 

Nevertheless, there are several caveats which we must mention for better estimating the robustness of these results. The first is that the near-Planckian regime occurs close to the bounce, a point where the scale factor is
nonvanishing. Therefore Eq.~\Eq{atp}, which was instrumental to the resolution of the perturbation equations, may not be a good approximation of the evolution of the universe during this first stage of expansion. A more complicated ansatz
for the scale factor may require numerical methods, which should be compared with the above analysis. A second issue is that we used perturbation theory in a regime which is intrinsically nonperturbative, so that the former may be in fact
invalid and deviation from scale invariance be a pathology of the linear approximation. A treatment beyond first-order perturbations will need to be developed to verify this apparent tension.

A scale-invariant or red-tilted tensor spectrum can be achieved in the interval $-1/q_\a< p\lesssim -1/(q_\a+1)$, where we took into account the threshold for having modes exiting rather than entering the horizon. It would be interesting to check whether these values also produce a scale-invariant scalar spectrum. Although a very small tensor-to-scalar ratio could accomodate a blue-tilted tensor index, the dependence on $r$ in Eq.~\Eq{ff} is logarithmic, so only a certain degree of fine tuning would allow for a large blue tilt. For instance, if $r\sim 10^{-8}$, these above bounds would still be $n_T\lesssim 1$. In order to make a safe estimate of Eq.~\Eq{ff} we should plug in the tensor-to-scalar ratio predicted by the model, which we do not yet know as we still miss the scalar spectrum. However, anomaly cancellation may not happen in the scalar sector in near-Planck regime, which may actually be a sign that perturbation theory fails to converge. In fact, anomaly cancellations have been shown only when inverse-volume corrections are very small ($\a=1+\dots$, where $(\dots)\ll 1$) \cite{bal}, that is, precisely in the large-volume, quasi-classical regime. Hence, for the time being the study of cosmological observables and consistency equations is not negatively affected by the eventual exclusion of the small-volume regime from the observable window. This will be subject of investigation in the near future.

{\it Note added:} Recently, tensor perturbations in the small-volume regime have been studied independently in Ref.~\cite{CMNS2}. Where our analyses overlap, they agree.



\begin{acknowledgments}
We thank A. Ashtekar, G. Nardelli and P. Singh for useful discussions, and are indebted to M. Bojowald for valuable comments and reading the manuscript. G.C. is supported by NSF grant PHY-0653127. G.H. thanks IGC, PSU where part of this work was done. G.H. is partially supported by NSERC of Canada.
\end{acknowledgments}


\end{document}